\documentclass[12pt]{iopart}
\usepackage{bbm}
\usepackage{graphicx}
\usepackage{dcolumn}
\usepackage{bm}
\usepackage{braket}
\expandafter\let\csname equation*\endcsname=\relax 
\expandafter\let\csname endequation*\endcsname=\relax 
\usepackage{amsmath,amssymb,amsthm} 
\usepackage{multirow}
\usepackage{graphicx}
\usepackage{comment}
\usepackage[colorinlistoftodos]{todonotes}
\usepackage[colorlinks=true, allcolors=blue]{hyperref}

\begin{document}

\title[Training QAOA without access to a QPU]{Training the Quantum Approximate Optimization Algorithm without access to a Quantum Processing Unit}
\author{Michael Streif$^{1,2}$, Martin Leib$^1$}
\address{$^1$Data:Lab, Volkswagen Group, Ungererstr. 69, 80805 M{\"u}nchen, Germany}
\address{$^2$University Erlangen-N{\"u}rnberg (FAU), Institute of Theoretical Physics, Staudtstr. 7, 91058 Erlangen, Germany}

\begin{abstract}
In this paper, we eliminate the classical outer learning loop of the Quantum Approximate Optimization Algorithm (QAOA) and present a strategy to find good parameters for QAOA based on topological arguments of the problem graph and tensor network techniques. Starting from the observation of the concentration of control parameters of QAOA, we find a way to classically infer parameters which scales polynomially in the number of qubits and exponentially with the depth of the circuit. Using this strategy, the quantum processing unit (QPU) is only needed to infer the final state of QAOA. This method paves the way for a variation-free version of QAOA and makes QAOA more practical for applications on NISQ devices. 
Moreover, we show the applicability of our method beyond the scope of QAOA, in improving schedules for quantum annealing.
\end{abstract}

\maketitle

%\tableofcontents

\section{Introduction}

The development of Shor's algorithm for integer factorization \cite{shor1994algorithms} and Grover's algorithm for searching an unstructured database \cite{grover1996fast}, with proven exponential and polynomial speed-up over their classical counterparts respectively, sparked the run on building first quantum processing units (QPUs) and culminated recently in first devices with up to tens of qubits. However, Shor's and Grover's algorithm require ten thousands of qubits \cite {jones2012layered} and error correction techniques. To achieve useful quantum computation already in the next decade it is necessary to develop algorithms which exploit the full power of these Noisy Intermediate Scale Quantum (NISQ) devices without relying on error correction codes.

Variational quantum algorithms, which are parameterized quantum circuits updated in classical learning loops, are seen as promising candidates. There exist many versions tailored for different fields of applications, such as the Variational Quantum Eigensolver (VQE) \cite{peruzzo2014variational} for finding the minimal energy state in quantum chemistry applications or Quantum Neural Networks (QNNs) \cite{mitarai2018quantum,killoran2018continuous} for quantum machine learning applications.  

The Quantum Approximate Optimization Algorithm (QAOA) \cite{farhi2014quantum} is a variational algorithm designed to solve combinatorial optimization problems. It was applied to NP-hard problems, such as Max-Cut \cite{farhi2014quantum}, Max-3-Lin-2 \cite{2014arXiv1412.6062F} or to sample from Boltzmann machines \cite{verdon2017quantum}. Moreover, it was shown that it is not possible to sample efficiently from the output state of QAOA with classical hardware \cite{farhi2016quantum} and that it is possible to achieve a Grover-type speed-up with QAOA \cite{jiang2017near}. However, due to its variational character, there exist only few insights in the performance and the scaling properties of QAOA in comparison to other methods, either of classical or quantum nature  \cite{zhou2018quantum, 2019arXiv190702359W, mbeng2019quantum, bapat2018bang, hastings2019classical, streif2019comparison}.

A major bottleneck of QAOA lies in the task of finding optimal parameters of the quantum circuit. Updating the parameters requires to estimate the energy expectation value of the output state and therefore repetitive QPU calls. Recently, different outer loop optimization strategies to mitigate this bottleneck were investigated, such as special-purpose learning strategies \cite{zhou2018quantum} or the use of black-box machine learning techniques \cite{verdon2019learning,wilson2019optimizing}. However, it is still unclear how these methods will perform in realistic setups. Thus, finding efficient ways to train quantum algorithms or to bypass the training will decide whether quantum variational algorithms allow for any quantum advantage. 

In the present contribution, we take a step into this direction and propose a novel method to infer the parameters of QAOA. Using this method, the QPU is not needed for updating the parameters, but only for sampling from the output state of QAOA. This eliminates an major obstacle of QAOA - its variational character. We show that this methods performs comparable or even better than the originally proposed version of QAOA, or vanilla QAOA, where we use outer loop learning optimization routines to train the variational algorithm. 

This paper is structured as follows. In Sec.~\ref{sec:theory}, we review QAOA and introduce the tree-QAOA strategy based on Tensor Network techniques. In Sec.~\ref{sec:applications}, we apply tree-QAOA to Max-Cut on 3-regular graphs and spin glasses on 2D grid lattices. Additionally, we compare the results with a vanilla QAOA setup where the parameters of the circuit are found by classical outer learning loops for each problem instance separately. In Sec.~\ref{sec:disorder}, we analyze the effect of disorder on the tree-QAOA results. In Sec.~\ref{sec:annealing}, we interpolate the found QAOA parameter to a schedule for quantum annealing and compare the performance to a linear schedule. Finally, in Sec.~\ref{sec:conclusion}, we conclude with a summary of our findings and an outlook.

\section{Theory}
\label{sec:theory}
\subsection{The Quantum Approximate Optimization Algorithm}
The landscape of quantum algorithms is to date divided into two areas: algorithms like the ones invented by Shor and Grover with provable speedup that however only can be executed on fully error corrected QPUs and quantum heuristics, without a proof of speedup, that are believed to optimally leverage the limited capabilities of the non-error corrected devices of the NISQ computing era. These heuristics are mainly variational quantum algorithms like VQE \cite{peruzzo2014variational}, QNNs \cite{biamonte2017quantum,farhi2018classification} and QAOA \cite{farhi2014quantum} which utilize parameterized gates where the parameters are optimized with classical computing resources. We focus on QAOA which is a variational ansatz to sample from solutions of combinatorial optimization problems.

Starting from a combinatorial optimization problem, the first step is to find a classical spin glass Hamiltonian with a ground state that represents its solution, and immediately promote it to its quantum version,
\begin{align}
\label{eq:probHam}
    \mathrm{H}_\mathrm{P}=\sum_{i,j}J_{ij}\sigma_z^{(i)} \sigma_z^{(j)}+\sum_i h_i \sigma_z^{(i)},
\end{align}
where every classical binary variable or spin $i$ is promoted to qubit $i$ with $\sigma_z^{(i)}$ its Pauli-Z operator. The $J_{ij}$ are interaction strengths between two qubits $i$ and $j$, and $h_i$ is the local energy offset for each qubit $i$. The spin glass can be represented by a graph $G_\mathrm{P}=(V,E)$ where each vertex $v\in V$ symbolizes a spin and vertices are connected by an edge $e\in E$  if the corresponding spins have a non-vanishing interaction, $J_{ij}\neq0$. Since finding the ground state of a spin glass is itself a NP-complete problem there exist mappings from every other combinatorial optimization problem in NP to a spin glass, many of them detailed in \cite{lucas2014ising}. With this we have transformed the problem of finding the solution to our combinatorial optimization problem to finding an assignment of the spins such that the energy of the spin glass is as low as possible. The next step in QAOA is to solve this problem with a variational ansatz, 
\begin{align}
    E_g = \min_{\left\{\beta_i, \gamma_i\right\}} \left\langle \Psi(\{\beta_i, \gamma_i\})| H_{\mathrm{P}}| \Psi(\{\beta_i, \gamma_i\})\right\rangle \,,
\end{align}
i.e. one searches for the optimal parameters $\{\beta_i^*, \gamma_i^*\}$ of the variational QAOA state $\left| \Psi(\{\beta_i, \gamma_i\})\right\rangle$ that minimize the expectation value $E_g$ of the problem Hamiltonian $H_\mathrm{P}$. 
Inspired by quantum annealing, the QAOA variational quantum state is generated by initializing the qubit register in an equal superposition of all computational basis states $\left|+\right\rangle = \bigotimes_{i= 1}^n(\left|0_i\right\rangle + \left| 1_i\right\rangle) / \sqrt{2}$ and is subsequently mapped with singly parametrized unitaries $U_{\mathrm{M}}(\gamma_i) = e^{-i\beta_i H_\mathrm{M}}$ and $U_{\mathrm{P}}(\gamma_i)=e^{-i\gamma_i H_\mathrm{P}}$ generated by the mixing Hamiltonian $H_\mathrm{M}=\sum_{i=1}^{n} \sigma_x^{(i)}$ with $\sigma_x^{(i)}$ the Pauli-X operator of qubit $i$ and the problem Hamiltonian $H_\mathrm{P}$, 
\begin{align}
    \left|\Psi(\{\beta_i, \gamma_i\})\right\rangle = U_\mathrm{M}(\beta_p)U_\mathrm{P}(\gamma_p) \dots U_\mathrm{M}(\beta_2)U_\mathrm{P}(\gamma_2)\underbrace{U_\mathrm{M}(\beta_1)U_\mathrm{P}(\gamma_1)}_{\text{QAOA block}} \left|+\right\rangle^{\otimes n}\,.
\end{align}
We call one application of the mixing combined with the problem unitary a QAOA block and define $p$ as the number of QAOA blocks. The optimal parameters $\{\beta_i^*, \gamma_i^*\}$ can be found by an outer classical learning loop that relies on repeated execution of the QAOA circuit on the QPU to estimate the value of $E_g$. In the present publication we show how to find optimal parameters without executing a single circuit on the QPU.
Once the optimal parameters are found, the QAOA state $\left| \Psi({\beta_i^*, \gamma_i^*})\right\rangle$ can be prepared and one can sample from low energy eigenstates of the problem Hamiltonian $H_\mathrm{P}$ by repeated measurements of the individual qubits in their $\sigma_z$-basis.

\subsection{Concentration of parameters}
\begin{figure}[t!]
\includegraphics[width=.75\columnwidth]{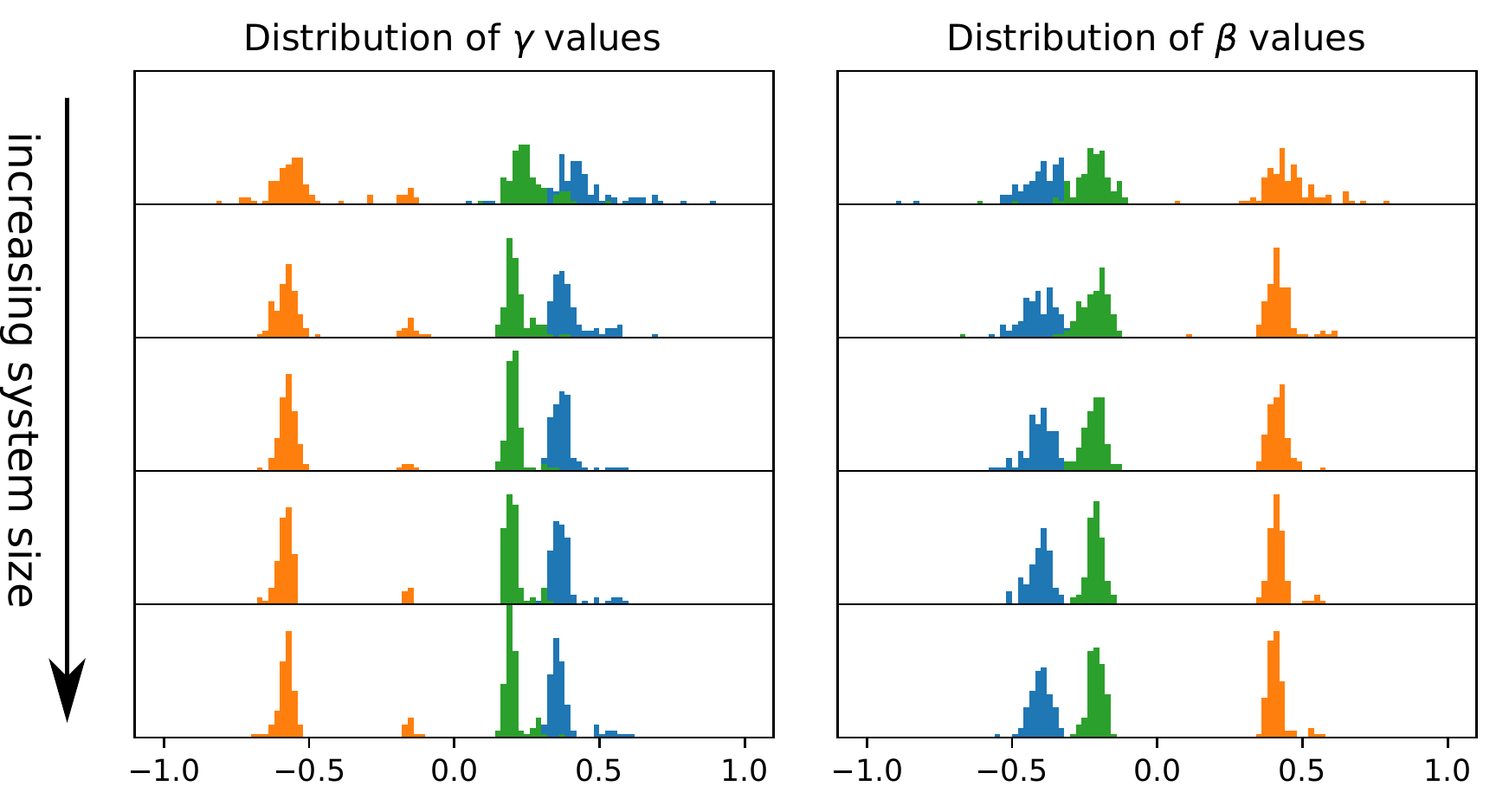}
\caption{Histograms of the optimal parameters when training a vanilla QAOA setup with $p=3$ QAOA blocks for 4-regular graphs with interaction strengths $J_{ij}=\{-1,1\}$. When increasing the system size, the distributions get narrower.}
\label{fig:concentration}
\end{figure}

To solve a specific instance of the combinatorial optimization problem with vanilla QAOA would involve numerous calls to the QPU in order to approximate $E_g$ to find the optimal QAOA parameters, followed by sampling from the QAOA output state. This means that finding the optimal parameters for the QAOA circuit would add to the computational complexity, or run time of QAOA itself. However it was quickly realized that finding the optimal parameters for each and every instance individually might not be necessary \cite{brandao2018fixed}. 
Consider an experiment where we randomly generate spin glass instances with coupling strengths drawn from $J_{ij} \in \{-1,0,1\}$ where every qubit is only coupled to a fixed amount of other qubits, i.e. $J_{ij}$ can only be nonzero for a fixed amount of times for fixed $i$. Subsequently we find the optimal parameters for every QAOA circuit generated by the specific instance individually and plot an estimate of the distribution of optimal parameters, cf. Fig.~\ref{fig:concentration}. The striking insight we get from this experiment is that the parameters are not equally distributed as one might expect but they rather concentrate around specific values. These specific values depend on the number of qubits every qubit interacts with however they do not depend on the size, i.e. the number of qubits involved, of the specific instance. Additionally the distribution of optimal parameters around the average value is getting narrower as the system size increases. These findings suggest that the optimal parameters do not strongly depend on the specific problem instance but rather the general topological features of the class of problems we are investigating. We might therefore find the average optimal parameters once and reuse them for all instances, thereby eliminating the computational cost per instance of finding the optimal parameters. In addition to that one can find the optimal parameters without any call to the QPU leveraging the experimental finding that the optimal parameters do not depend on the problem size. 

One might consider sampling random instances of the problem class that are small enough to simulate classically and find the optimal parameters for them \cite{brandao2018fixed}. For small instances however the variance of optimal parameters is bigger than for the large instances, therefore one has to average over a large number of instances to get a good estimate of the average optimal parameters. In this publication we show how to classically calculate $E_g$, and find the optimal parameters in the opposite limit: for an instance that is formally of infinite size. In addition to getting rid of the averaging problem we numerically show that the optimization landscape for the infinite size instance is much easier to navigate with the standard gradient based optimizer routines than the optimization landscape of the instances that are small enough to be simulated classically. 
 
In order to understand why it is possible to simulate QAOA instances of formally infinite size it is necessary to introduce the notion of the \textit{reverse causal cone} which was also mentioned in the original QAOA paper \cite{farhi2014quantum}. $E_g$, the expectation value of the problem Hamiltonian for the QAOA state can be decomposed into a sum of expectation values of correlation functions between qubits that interact,
\begin{align}
    E_g = \sum_{\left( i,j\right)\in E} J_{ij} \left\langle  \Psi(\{\beta_i, \gamma_i\})\right| \sigma_z^{(i)}\sigma_z^{(j)} \left|\Psi(\{\beta_i, \gamma_i\})\right\rangle \,.
\end{align}
The unitaries $U_{\mathrm{M}}$ and $U_{\mathrm{P}}$ that form the QAOA state can be decomposed into one-qubit $\sigma_x$-gates and two-qubit $\sigma_z\sigma_z$-gates,
\begin{align}
    U_{\mathrm{M}}(\beta)&= \exp(-i\beta\sum\limits_i \sigma_x^{(i)}) = \prod\limits_i \exp(-i\beta \sigma_x^{(i)})= \prod\limits_i U_{\mathrm{M}}^{(i)}(\beta),\\
    U_{\mathrm{P}}(\gamma)&= \exp(-i\gamma\sum_{\left( i,j\right)\in E} \sigma_z^{(i)}\sigma_z^{(j)}) = \prod_{\left( i,j\right)\in E} \exp(-i\gamma \sigma_z^{(i)}\sigma_z^{(j)})= \prod_{\left( i,j\right)\in E} U_{\mathrm{P}}^{(i,j)}(\gamma)\,.
\end{align}
With this decomposition we can considerably simplify the individual correlation functions that make up the expectation value $E_g$. We start by commuting the one-qubit gates of the last application of the mixing Hamiltonian with the $\sigma_z$-operators of the correlation function between qubit $i$ and $j$ for all qubits but qubits $i$ and $j$,
\begin{align}
     &\left\langle  \Psi(\{\beta_i, \gamma_i\})\right| \sigma_z^{(i)}\sigma_z^{(j)} \left|\Psi(\{\beta_i, \gamma_i\})\right\rangle = \nonumber\\
     &\left\langle + \right|U_{\mathrm{P}}^\dag(\gamma_1)\dots U_{\mathrm{P}}(\gamma_p)\underbrace{\left(U_{\mathrm{M}}^{(i)\dag}(\beta_p)\sigma_z^{(i)}U_{\mathrm{M}}^{(i)}(\beta_p)\right)\left(U_{\mathrm{M}}^{(j)\dag}(\beta_p)\sigma_z^{(j)}U_{\mathrm{M}}^{(j)}(\beta_p)\right)}_{\hat{O}(\{i,j\})} U_{\mathrm{P}}(\gamma_p) \dots U_{\mathrm{P}}(\gamma_1)\left| + \right\rangle\,,
\end{align}
where we introduced the placeholder $\hat{O}(M)$ for an operator with support given by the tensor product of the Hilbert space of the qubits in the set $M$.
This mathematical reformulation employing the unitarity of quantum gates reflects the physical principle that the one-qubit gates in the last layer of the QAOA circuit on qubits that are not qubits $i$ and $j$ do not affect the correlation function between qubits $i$ and $j$, i.e. they are not within their \textit{reverse causal cone}. To evaluate the full \textit{reverse causal cone} we have to iterate this procedure through the entire quantum circuit. The next step is to commute every two-qubit $\sigma_z\sigma_z$-gate that does not involve either qubit $i$ nor qubit $j$ of the last application of the problem Hamiltonian,
\begin{align}
    &\left\langle + \right|U_{\mathrm{P}}^\dag(\gamma_1)\dots U_{\mathrm{P}}^\dag(\gamma_p)\hat{O}({i,j}) U_{\mathrm{P}}(\gamma_p) \dots U_{\mathrm{P}}(\gamma_1)\left| + \right\rangle = \nonumber\\
    &\left\langle + \right|U_{\mathrm{P}}^\dag(\gamma_1)\dots \underbrace{\prod\limits_{k,l \in N(\{i,j\})} {U_{\mathrm{P}}^{(k,l)}}^\dag(\gamma_p)\hat{O}(\{i,j\}) \prod\limits_{k,l \in N(\{i,j\})}U_{\mathrm{P}}^{(k,l)}(\gamma_p)}_{\hat{O}(N(\{i,j\}))} \dots U_{\mathrm{P}}(\gamma_1)\left| + \right\rangle\,.
\end{align}
The application of the problem Hamiltonian enhances the support of the placeholder operator in the middle of the expectation value to encompass the Hilbert space of the qubits in the set  $N(\{i,j\}) = \{i,j\} \, \cup \, \{k | \text{qubit k interacts with either qubit i or qubit j\}}$ that includes qubits $i$ and $j$ as well as all qubits that interact with qubits $i$ and $j$. If we recursively progress through the layers with this process we get the set of gates that influence the correlation function between qubit $i$ and $j$. We call this set of gates the \textit{reverse causal cone}. 

If we think of the graph representing the classical spin glass, the \textit{reverse causal cone} has an intuitive visualization: The $\sigma_z^{(i)}\sigma_z^{(j)}$ operator is symbolized by the edge between vertex $i$ and $j$ and we can construct the support of the \textit{reverse causal cone} by recursively adding all neighbors of the vertices that are already in the set starting with vertices $i$ and $j$. Once we repeated this process p times we have found the support of the \textit{reverse causal cone}. We can therefore think of the support of the \textit{reverse causal cone} as a subgraph generated with the above explained recipe of the full problem graph that defines the classical spin glass. We denote the state generated by the application of the \textit{reverse causal cone} as $\ket{\mathrm{RCC}_\mathrm{subgraph}}$, where the subgraph represents the support of the \textit{reverse causal cone}. 

We consider the  problem instance large compared to the number of blocks if the subgraphs induced by the correlation functions of $E_g$ are proper subgraphs of the problem graph. For random problem graphs with fixed degree and $J_{ij}=\{-1,0,1\}$ that are large compared to the number of blocks the most likely subgraph is a tree with degree given by the degree of the problem graph. In the limiting case of an infinite size problem graph we can therefore approximate $E_g$ with,
\begin{align}
  e_g \equiv\lim_{|V(G_{\mathrm{P}})|\to \infty}\frac{E_g}{|E(G_{\mathrm{P}})|} \to  \left\langle \text{RCC}_{\mathrm{Tree}}\right| \sigma_z^{(1)}\sigma_z^{(2)} \left|\text{RCC}_{\mathrm{Tree}}\right\rangle \,,
\end{align}
where $|V(G_{\mathrm{P}})|$ and $|E(G_{\mathrm{P}})|$ are the number of vertices and edges of the problem graph respectively and the right hand side is the correlation function of generic qubits $1$ and $2$ of the \textit{reverse causal cone} defined by a tree with degree given by the degree of the problem graphs. In the following section we are going to present a method based on tensor networks to calculate this specific correlation function classically efficient. 

\subsection{Tree-QAOA}
In this section we introduce the tree-QAOA, a method that eliminates the need to infer $E_g$ from repeated execution of the QAOA circuit on the QPU, for which the QPU is only required to sample from the final state of the quantum algorithm.
Our calculation of $e_g$ for an infinite size instance relies on methods from the theory of tensor networks. To make this article self consistent we will provide a short introduction to tensor networks from the perspective of simulating quantum circuits and afterwards introduce the tree-QAOA method. 

Tensor networks are a very successful technique for the resource efficient representation of states of quantum many-body systems. They are used for numerical calculations and their powerful graphical representation and calculus can be used for exact proofs \cite{orus2014practical}. 

\begin{figure}
\includegraphics[width=1\columnwidth]{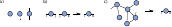}
\caption{(a) A tensor representing a scalar, a vector and a matrix. (b) A tensor representation of a standard matrix product $\sum_j A_{i,j}B_{j,k}$. (c) A tensor network with two dangling bonds, which results in a rank-2-tensor after the contraction.}
\label{fig:tensors}
\end{figure}
The basic building block of tensor networks are rank $r$ tensors or $r$-dimensional matrices. Tensors are generalizations of scalars, vectors and matrices to higher dimensions, i.e. a scalar $s$ is a rank-0-tensor, a vector $v_i$ is a rank-1-tensor and a matrix $M_{ij}$ a rank-2-tensor, see Fig.~\ref{fig:tensors}(a). A tensor of rank $r$ can be symbolized by a vertex with $r$ incident edges, also called bonds. Tensors also support basic matrix operations like the standard matrix product between matrix $A$ and $B$, cf. Fig.~\ref{fig:tensors}(b), and generalizations thereof, 
\begin{align}
    T_{i,k} &= \sum\limits_j A_{i,j} B_{j,k}, & T_{i_1, \dots ,i_n, k_1, \dots ,k_m} = \sum\limits_{j_1, \dots j_l} A_{i_1, \dots , i_n, j_1\dots ,j_l} B_{j_1, \dots, j_l,k_1,\dots, k_m}\,.
\end{align}
The generalization of the matrix product is called tensor contraction and has a nice graphical representation: If one wants to symbolize the operation of summing two tensors over a common index one connects the two vertices with the bond corresponding to the specific index. Tensor networks are a further generalization of all the aforementioned: they are a collection of tensors connected with specific contraction rules and can be symbolized by a graph where all contractions are symbolized by bonds connecting the vertices, i.e. tensors, in the graph. Not all bonds in this graph need to be incident to two vertices, some of them might only be connected to one vertex thus defining a dangling bond, cf. Fig~\ref{fig:tensors}(c). Every tensor network is the representation of a tensor of rank given by the number of dangling bonds, because every non-dangling bond is to be contracted, i.e. summed over. Without loss of generality we assume the number of distinct values of the indices, or bond dimension, to be fixed and equal for all indices. However we specifically include the possibility of multiple bonds between every pair of tensors.

A quantum circuit is a sequence of gates that are applied to a specified initial quantum state. To calculate the outcome of a quantum circuit one needs to find an appropriate representation of gates as matrices and states as vectors in the Hilbert space of the qubit register and multiply the gates with the initial state in the order prescribed by the quantum circuit. This is why a quantum circuit can also be described as a tensor network where the tensors are the matrix representations of the gates and the bonds are the qubit lines connecting the temporal sequence of gates that act on a specific qubit. We are interested in directly calculating the mean value of operator $\sigma_z^{(1)}\sigma_z^{(2)}$ for the state $\left|\text{RCC}_{\mathrm{Tree}}\right\rangle$. Because the resulting tensor network is smaller we decided to follow the approach by \cite{fried2018qtorch} and reformulate our mean value in terms of density matrices,
\begin{align}
\label{eq:TreeRCC}
    \left\langle \text{RCC}_{\mathrm{Tree}} \right| \sigma_z^{(1)}\sigma_z^{(2)}\left|\text{RCC}_{\mathrm{Tree}}\right\rangle = \text{tr}\left[\sigma_z^{(1)}\sigma_z^{(2)} \left| \text{RCC}_{\mathrm{Tree}} \right\rangle \left\langle\text{RCC}_{\mathrm{Tree}}\right|\right]
\end{align}
The application of a gate in the density matrix representation can be reformulated again to a matrix product version with vectorized density matrix,
\begin{align}
    U^\dag \rho U \qquad \to \qquad \left(U^\dag \otimes U\right) \vec{\rho}\,.
\end{align}
We use this superoperator picture in analogy to \cite{markov2008simulating} to get the tensor network representing the mean value Eq.~(\ref{eq:TreeRCC}) with the following rules: Every gate in the tensor network of the QAOA circuit is represented by the superoperator version of the gate matrix in the computational basis. The dangling bonds for the initial state of the QAOA circuit are connected to rank-1 tensors, $t = (0.5, 0.5, 0.5, 0.5) \widehat{=} |+\rangle\langle + |$, i.e. vectorized versions of the projector on the state $|+\rangle$, while the dangling bonds at the end of the circuit are connected to either $M_z = (1, 0, 0, -1)$, the vectorized version of the measurement operator for $\sigma_z$, for qubits $1$ and $2$ or $M_{\text{tr}}= (1,0,0,1)$ for all remaining qubits. The resulting tensor network does not have any dangling bonds anymore and its contraction provides us with the value of $e_g$, c.f. Eq.~(\ref{eq:TreeRCC}).

Tensor networks of quantum circuits can be very memory efficient ways to save quantum states. However, to evaluate mean values with respect to these states, the tensor networks need to be contracted, which might result in multiplication of very large matrices. The number of elementary operations we have to execute, or computational complexity, of contracting a single bond between two tensors of rank $r_1$ and $r_2$ respectively is given by $d^{r_1 + r_2 + 1}$, where $d$ is the bond dimension. In our case as detailed above $d=4$ because we work in the superoperator picture. Contracting a tensor network consists of deciding on a contraction sequence, i.e. a sequence of bonds that are to be contracted one after the other, followed by the actual contraction. Therefore the computational complexity of contracting a tensor network is given by the sum of the elementary operations of the individual contractions. Contracting the entire tensor network is dominated by the largest computational complexity of contracting a single bond, i.e. the contraction involving the pair of tensors with the largest sum of ranks. The computational complexity of a tensor network contraction does strongly depend on the chosen contraction sequence. However it has been shown that the computational complexity of the optimal contraction sequence scales with the exponential of the treewidth of the graph underlying the tensor network \cite{markov2008simulating}. The treewidth is a positive integer that can be assigned to every graph that intuitively is a measure of how close the graph is to a tree. The treewidth is defined as the minimal width of all possible tree decompositions. A tree decomposition of a graph G is a tree, T, with vertices $V_1, \dots , V_n$ where each vertex is a subset of the vertices of the original Graph G. The subsets $V_i$ have to fulfill the following constraints: Every graph vertex has to be in at least one set $V_i$. The sets $V_{i_1}, \dots, V_{i_k}$ that contain vertex $i$ of the original graph form a connected subtree. For every edge $(i, j)$ in the original graph, there is at least one subset $V_i$ that contains both $i$ and $j$. Consequently the treewidth of a tree is 1 and the treewidth of the fully connected graph is the number of vertices minus 1. 

In this section, we detail a specific, not necessarily optimal, contraction scheme for the tensor network representing an arbitrary QAOA correlation function defined by its \textit{reverse causal cone}. With this we can show that the scaling of the contraction complexity to evaluate the QAOA correlation function is upper bounded by an exponential function of the number of blocks in the QAOA circuit times the treewidth of the subgraph induced by the \textit{reverse causal cone} of the correlation function.

\begin{center}
\begin{figure*}[t!]
\centering
\includegraphics[width=1\textwidth]{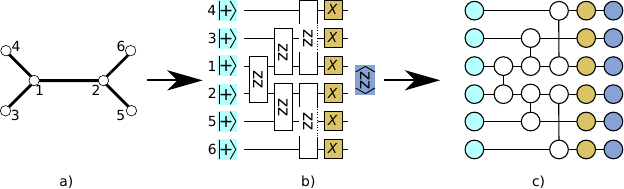}
\begin{minipage}[t!]{1\linewidth}
	\caption{An example how to generate the QAOA circuit and its tensor network representation from an initial problem graph. (a) A (binary) tree graph of degree 3. White circles represent vertices, black lines represent connections between vertices. (b) The resulting tree-QAOA circuit for $p=1$. All qubits are initialized in the $\ket{+}$ state. (c) The tensor network representing the tree-QAOA circuit. Initially the maximal rank of a tensor in the network is three.}
	\label{fig:4local}
\end{minipage}
\end{figure*}
\end{center}

We employ a contraction scheme similar to the one in \cite{markov2008simulating} for the circuits for one-way quantum computation. The first step is to decompose the two qubit $\sigma_z\sigma_z$-gates which are rank-4 tensors into to rank-3 tensors, c.f. Fig.~\ref{fig:4local}. The two qubit $\sigma_z\sigma_z$-gate can be written as,
\begin{align}
    \exp(-i\gamma \sigma_z^{(i)}\sigma_z^{(j)}) = \left|0_i\right\rangle\left\langle 0_i\right| \otimes \exp(-i\gamma\sigma_z^{(j)}) + \left|1_i\right\rangle\left\langle 1_i\right| \otimes \exp(i\gamma\sigma_z^{(j)})\,,
\end{align}
which suggests the decomposition of this gate into two rank three tensors $a_{k,l,m}$ and $b_{n,o,p}$ where the first two indices of both tensors are the physical indices of the Hilbert space of the individual qubits and we additionally introduced a virtual bond that is to be summed over to get the full gate,
\begin{align}
    a_{\_,\_,0}&= \left|0_i\right\rangle\left\langle 0_i\right| &
    a_{\_,\_,1}&= \left|1_i\right\rangle\left\langle 1_i\right| \\
    a_{\_,\_,0}&= \exp(-i\gamma\sigma_z^{(j)}) &
    a_{\_,\_,1}&= \exp(i\gamma\sigma_z^{(j)}) \,.
\end{align}
For the superoperator version of the gate we proceed in complete analogy to get, 
\begin{align}
    \exp(-i\gamma \sigma_z^{(i)}\sigma_z^{(j)}) \to &\exp(-i\gamma \sigma_z^{(i)}\sigma_z^{(j)}) \otimes \exp(i\gamma \sigma_z^{(i)}\sigma_z^{(j)}) = \nonumber\\
    &\left|0_{i,l}0_{i,r}\right\rangle\left\langle 0_{i,l}0_{i,r}\right| \otimes \exp(-i\gamma\sigma_z^{(j,r)})\otimes \exp(i\gamma\sigma_z^{(j,l)}) + \nonumber\\ &\left|1_{i,l}1_{i,r}\right\rangle\left\langle 1_{i,l}1_{i,r}\right| \otimes \exp(i\gamma\sigma_z^{(j,r)})\otimes \exp(-i\gamma\sigma_z^{(j,l)}) + \nonumber\\
    &\left|0_{i,l}1_{i,r}\right\rangle\left\langle 0_{i,l}1_{i,r}\right| \otimes \exp(-i\gamma\sigma_z^{(j,r)})\otimes \exp(-i\gamma\sigma_z^{(j,l)}) + \nonumber\\
    &\left|1_{i,l}0_{i,r}\right\rangle\left\langle 1_{i,l}0_{i,r}\right| \otimes \exp(i\gamma\sigma_z^{(j,r)})\otimes \exp(i\gamma\sigma_z^{(j,l)}) \,.
\end{align}
From this decomposition we can directly see that the dimension of the virtual bond is 4 for the superoperator version of the two qubit $\sigma_z\sigma_z$-gate, i.e. the same dimension as the other bonds that connect gates that act on the same qubit. 

Our proposed contraction sequence starts, after the above decomposition of the 2-qubit gates, by contracting along the qubit lines. This leaves us with a tensor network that is isomorphic to the subgraph generated by the \textit{reverse causal cone}. The multiplicity of the bonds in the remaining tensor network however is in the maximal case given by the number of blocks in the QAOA circuit. The contraction complexity of the remaining tensor network therefore scales with the exponential of the treewidth of the subgraph multiplied by the number of QAOA blocks.

In our case of $e_g$, the treewidth of the generated subgraph is 1 and the tensor network calculation therefore scales exponentially only in the number of QAOA blocks but not in the number of qubits any more. We note that this result is only an upper bound. In fact, in Sec.~\ref{sec:applications}, we use a heuristic to find a good contraction ordering which does not exhaust this upper bound. 

\section{Numerical results}
\label{sec:applications}
In this section, we apply our method to two problems: Max-Cut problems on 3-regular graphs and spin glasses on square grids. Both of these problems possess individual characteristics: the Max-Cut problems on regular graphs exactly fit to the assumptions we made for tree-QAOA, whereas the spin glasses with non-fixed degree enables us to benchmark how tree-QAOA performs in situations where our assumptions are not perfectly satisfied.

To study the performance of tree-QAOA, we compare its outcome with a QAOA setup where we train the variational parameters of the quantum circuit for each instance separately by optimizing the expectation value $\braket{\Psi(\{\beta_i,\gamma_i\})|H_\mathrm{P}|\Psi(\{\beta_i,\gamma_i\})}$. To find good parameters, $\{\gamma_i^*,\beta_i^*\}$, we use two classical optimization routines, scipys implementation of L-BFGS-B (Broyden-Fletcher-Goldfarb-Shanno) \cite{liu1989limited} and Adam (Adaptive Moment Estimation) \cite{kingma2014adam}. To distinguish between the methods, we refer to them as tree-QAOA, BFGS-QAOA and Adam-QAOA. As a figure of merit we use the residual energy
\begin{align}
    \label{eq:ratio}
    r=\frac{\braket{\Psi(\{\beta_i,\gamma_i\})|H_\mathrm{P}|\Psi(\{\beta_i,\gamma_i\})}-E_0}{E_\mathrm{max}-E_0} \in [0,1],
\end{align}
which is a measure of how close the energy of the final output state is to the energy of the ground state. In this expression, $E_0$ denotes the ground state energy of the problem Hamiltonian $H_\mathrm{P}$ and $E_\mathrm{max}$ the energy of the highest excited state. We note that we also used other metrics, such as the overlap of the final state with the ground state(s), which yield qualitatively similar results as the ones presented in the following. However, the residual energy embodies a better metric for situations where one is interested in any good solution rather then only the optimal solution.

\subsection{Max-Cut on 3-regular graphs}
\label{sec:maxcut}
Max-Cut is the task to find a bipartition of the set of vertices $V$ of a graph $G=(V,E)$ such that the number of edges $E$ connecting vertices from one bipartition to the other is maximized. Max-Cut is known to be NP-hard and serves as reference in many complexity theory studies as well as for an initial benchmark of QAOA \cite{farhi2014quantum}. The spin glass Hamiltonian representing Max-Cut includes two-body interaction terms between all vertices in the edge set $E$,
\begin{align}
    \label{eq:maxcutspinglass}
    H_\mathrm{MaxCut}=\sum_{\left( i,j\right)\in E}\frac{1}{2}\left(\mathbf{1}+\sigma_z^{(i)}\sigma_z^{(j)}\right).
\end{align}
Finding the ground state of the spin glass is equivalent to solving the Max-Cut problem. In this section, we solely focus on Max-Cut problems on 3-regular graphs, i.e. graphs where each vertex has exactly 3 connections.

To find the control parameters of QAOA with tree-QAOA, we simulate QAOA on the tree subgraph corresponding to the problem graph. For 3-regular graphs, this is a tree with degree 3, i.e. a binary tree. For this tree structure, the number of qubits in the tree-QAOA circuit grows with the number of tree-QAOA blocks $p$ as $2^{p+2}-2$.  For this numerical experiment, we calculate the tree-QAOA parameters $\{\Vec{\gamma}^*_\mathrm{tree},\Vec{\beta}^*_\mathrm{tree}\}$ up to $p=8$ QAOA blocks, for which the circuit includes $N=1022$ qubits. To find a good contraction sequence of the tensor network that represents the tree-QAOA circuit we use the random-greedy method of the Python package opt einsum \cite{smith2018opt} and the L-BFGS-B algorithm to update its parameters. For $p>1$, we use the found parameters of $p-1$ blocks together new parameters which we add such that both $\{\gamma_i\}$ and $\{\beta_i\}$ form a linear schedule as initial guess for the optimization. We note that other optimization routines, hyperparameter tuning or more sophisticated contraction orderings could speed up the calculation. We note that we run the code on a standard off-the shelf desktop computer. 

\begin{figure*}[t!]
    \centering
    \begin{minipage}{0.5\textwidth}
        \centering
        \includegraphics[width=1\textwidth]{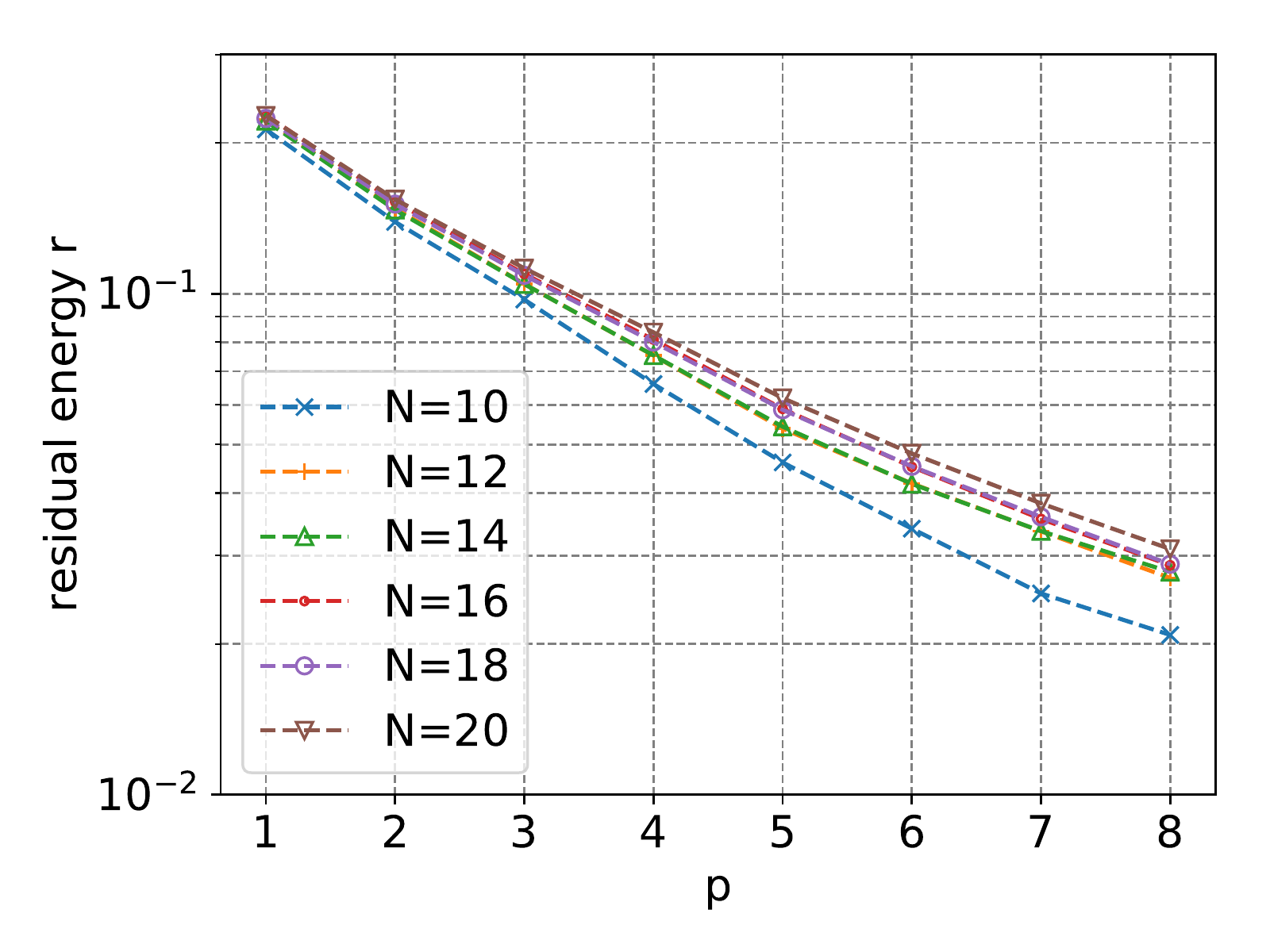}
    \end{minipage}\hfill
    \begin{minipage}{0.5\textwidth}
        \centering
        \includegraphics[width=1\textwidth]{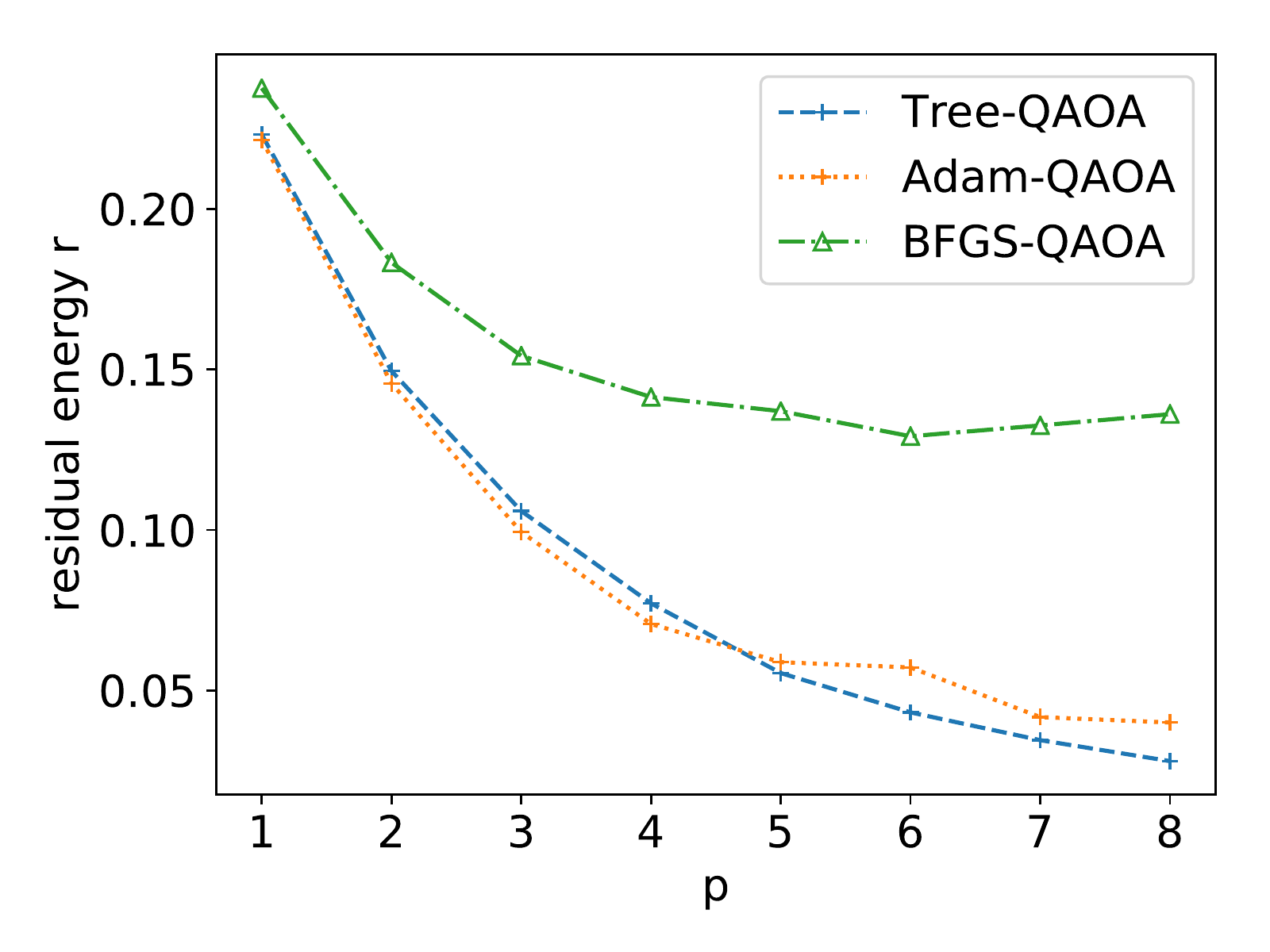}
    \end{minipage}
\caption{(a) The residual energy $r$ as defined in Eq.~(\ref{eq:ratio}) obtained with tree-QAOA averaged over $M=100$ Max-Cut problems on 3-regular instances for various system sizes $N=\{10,12,14,16,18,20\}$ in dependence of the number of QAOA blocks $p$. (b) A comparison of the averaged residual energy between tree-QAOA, BFGS-QAOA and Adam-QAOA for the same $12$-spin instances as used in (a).}
\label{fig:maxcut}
\end{figure*}
The optimal tree-QAOA parameters are subsequently used in QAOA for each Max-Cut instance. We randomly generate $M=100$ instances and calculate the average residual energy $r$, cf. Eq.~(\ref{eq:ratio}), for tree-QAOA, BFGS-QAOA and Adam-QAOA. For each instance, we run both classical optimization once with random initial parameter guesses. In Fig.~\ref{fig:maxcut}(a), we plot the  residual energy averaged over all instances using tree-QAOA for various system sizes as a function of the number of QAOA blocks $p$. We find a similar exponential scaling as observed by \cite{zhou2018quantum}. In Fig.~\ref{fig:maxcut}(b), we show a comparison between tree-QAOA, BFGS-QAOA and Adam-QAOA for $N=12$ spins. Notably, tree-QAOA performs comparable or even better than both of its competitors. For a small number of QAOA blocks, Adam-QAOA is slightly better, while for $p>5$ tree-QAOA is superior. BFGS-QAOA performs poorly in comparison to the other methods. We note that a comparison of classical optimization routines, as in \cite{wilson2019optimizing}, is not the scope of this paper and that it might be possible to find better residual energies by tuning the hyper-parameters, such as using multiple initial starting points or stronger convergence criteria. Using special-purpose methods for initial parameters guesses  \cite{zhou2018quantum} also showed improvement over off-the shelf optimizers but requires numerous repetitions of the quantum circuit. We note that while optimizing both the tree-QAOA and vanilla QAOA with L-BFGS-B with default settings, we observed that it was easier to find good parameters for tree-QAOA, which indicates that the parameter landscape for tree-QAOA is less rugged than for vanilla QAOA.

\begin{figure*}[t!]
    \centering
    \begin{minipage}{0.5\textwidth}
        \centering
        \includegraphics[width=1\textwidth]{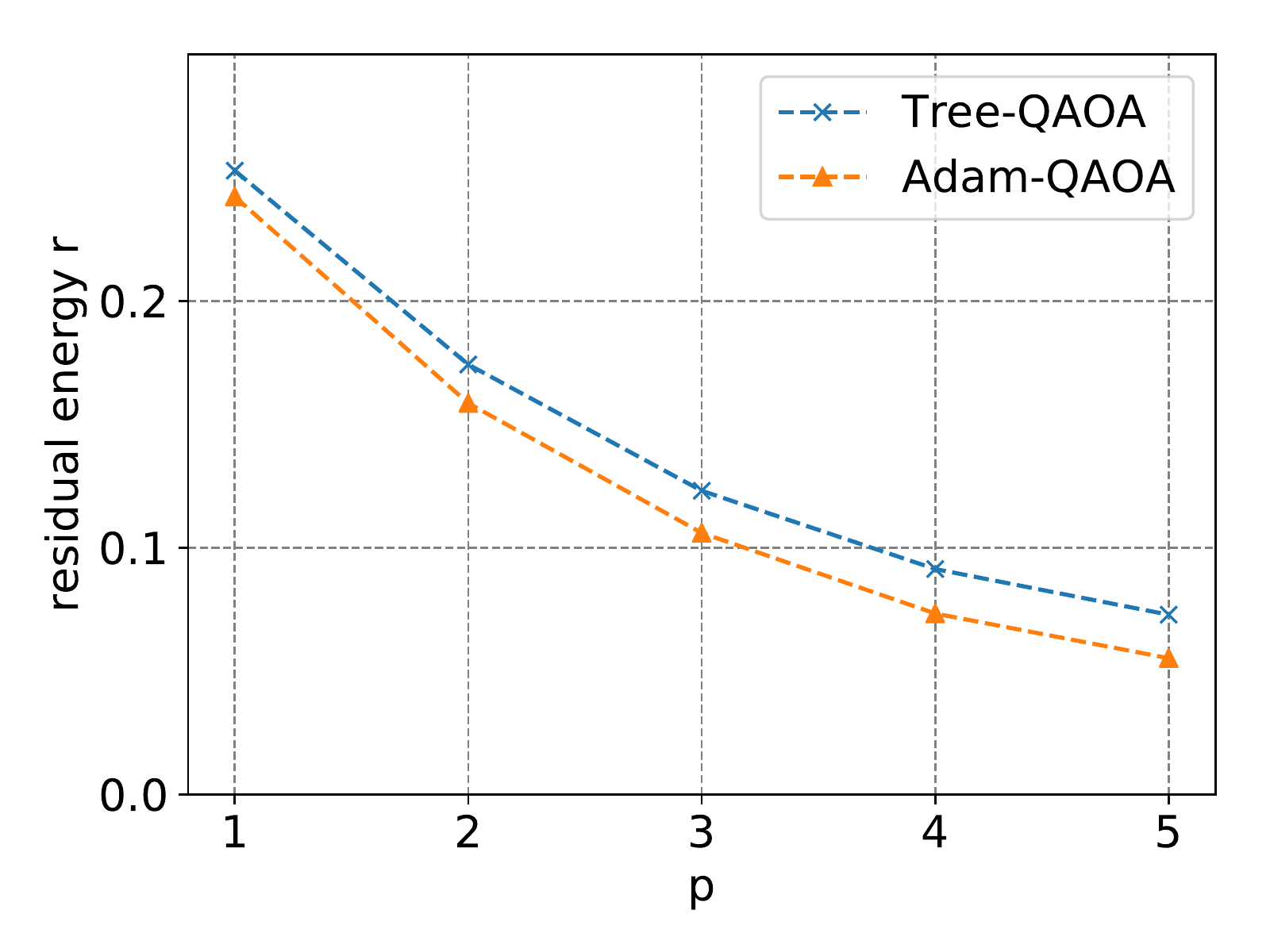}
    \end{minipage}\hfill
    \begin{minipage}{0.5\textwidth}
        \centering
        \includegraphics[width=1\textwidth]{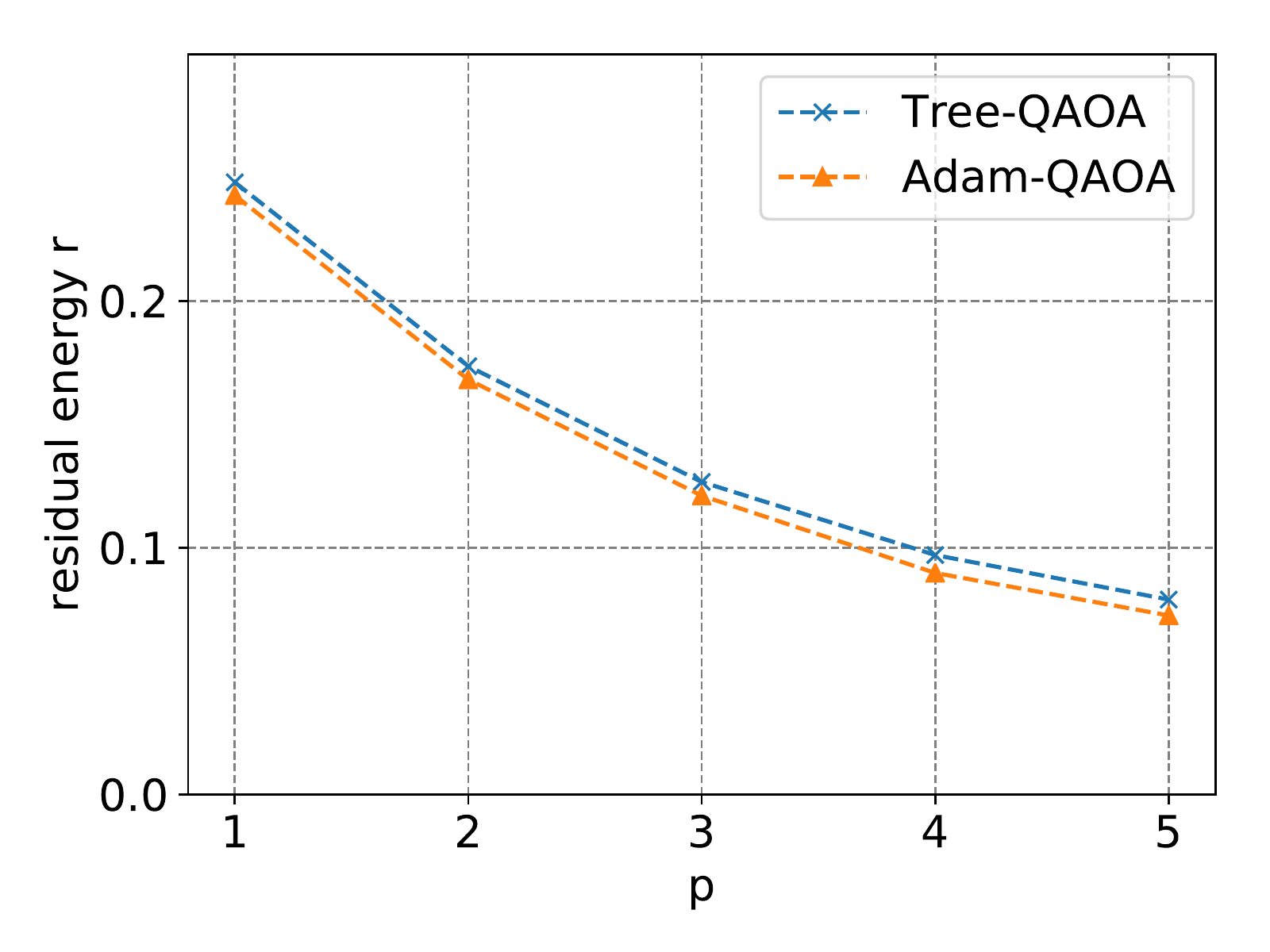}
    \end{minipage}
\caption{The residual energy $r$ averaged over $M=100$ 2D spin glasses obtained with tree-QAOA and Adam-QAOA for (a) a $3\times3$ square grid  and (b) a $4\times4$ square grid.}
\label{fig:weightedmaxcut}
\end{figure*}

\subsection{Spin glasses on 2D lattices}

In this section we apply the tree-QAOA method to spin glasses on two square grid structures, a $3\times3$ grid including 9 spins and $4\times 4$ including 16 spins.  We note that spin glass systems without local fields on 2D structures are classically solvable in polynomial time \cite{barahona1982computational}. However, this problem class helps us to understand how our method performs for a graph with non-regular degree. In the bulk, where $N\rightarrow\infty$, the average degree of the graph converges to 4. However, for the here studied $3\times3$ and $4\times 4$ grids, only 1, respectively 4 vertices have a degree of 4.  Moreover, the qubits of most current QPUs are arranged in 2D lattices, which makes this problem a natural fit which could be implemented without using swap gates. Therefore this setup possess the possibility to benchmark QAOA with the largest possible $p$ in a real experiment. The Hamiltonian describing the spin glass reads
\begin{align}
    H_\mathrm{2D}=-\sum_{\left( i,j\right)\in E} J_{ij} \sigma_z^{(i)} \sigma_z^{(j)},
\end{align}
where $E$ defines the edge set of the square lattices and coupling strengths are randomly drawn from an uniform distribution with $J_{ij}=\pm 1 \phantom{.}\forall i,j$.  As the average degree of the vertices of the graph tends to 4 for $N\rightarrow\infty$, we calculate the tree-QAOA parameters for a tree of degree 4. For this case, the tree-QAOA circuit includes $3^{p+1}-1$ qubits. In Fig.~\ref{fig:weightedmaxcut}(b) we plot the residual energy obtained with tree-QAOA for $N=\{9,16\}$ spin instances in comparison to Adam-QAOA. For small system sizes, tree-QAOA performs worse than Adam-QAOA. For larger system sizes the performance of both methods is comparable. This can be understood as the average degree of the graph is closer to 4. Surprisingly, also for the smaller system size, where our assumption that most subgraphs have a degree of 4 is very inaccurate, the outcome of tree-QAOA is much better than random guessing and shows a monotonically decreasing residual energy.

\section{Performance under the influence of disorder}
\label{sec:disorder}

\begin{figure*}[t!]
        \centering
        \includegraphics[width=0.5\textwidth]{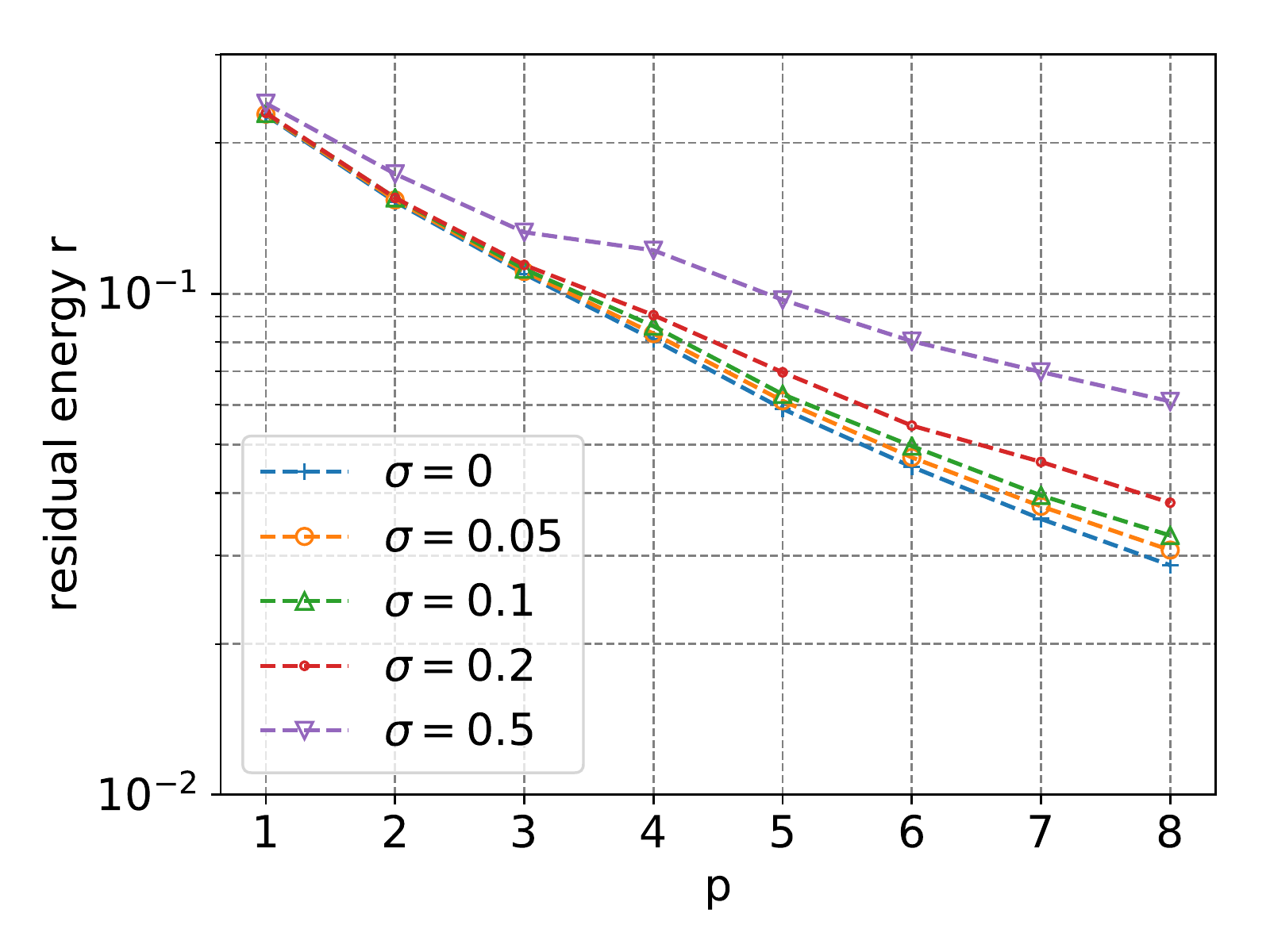}
\caption{The influence of different disorder levels on tree-QAOA for Max-Cut problems on 3-regular graphs with $N=16$ vertices. The disorder is drawn from a normal distribution with mean $\mu=0$ and standard deviation $\sigma$.}
\label{fig:disorder}
\end{figure*}
In realistic experiments, the performance of quantum algorithms will be influenced by analog control errors. Especially for variational algorithms, where the parameter optimization relies on the accurate evaluation of the loss function, such effects can corrupt the result. In this section we show that tree-QAOA is robust against a considerable amount of analog control errors.

To do so, we reuse the Max-Cut instances on 3-regular graphs with $N=16$ qubits studied in Sec.~\ref{sec:maxcut}. We consider analog control errors to result in parameterized gates that either over- or under-rotate by random amounts. We model this effect by adjusting the mixing and problem Hamiltonian in the following way,
\begin{align}
    H_\mathrm{MaxCut}\rightarrow \tilde{H}_\mathrm{MaxCut}&=\sum_{\left( i,j\right)\in E}\frac{1}{2}\left(\mathbf{1}+(1+\Delta)\sigma_z^{(i)}\sigma_z^{(j)}\right),\\
    H_\mathrm{M}\rightarrow \tilde{H}_\mathrm{M}&=\sum_i (1+\Delta) \sigma_x^{(i)},
\end{align}
where $\Delta\sim\mathcal{N}(0,\sigma)$. The amount of disorder can be controlled by the standard deviation of the  normal distribution, $\sigma$. We then simulate the QAOA circuit under the disturbed Hamiltonians, $\tilde{H}_\mathrm{MaxCut}$ and $\tilde{H}_\mathrm{M}$, while taking the expectation value  $\braket{\Psi(\{\beta_i,\gamma_i\})|H_\mathrm{MaxCut}|\Psi(\{\beta_i,\gamma_i\})}$, with respect to the undisturbed Hamiltonian, $H_\mathrm{MaxCut}$. This approach simulates the situation in a real experiment, where the disturbed Hamiltonian is unknown and the energy has to be estimated by the sum of all Pauli terms of the undisturbed Hamiltonian. In Fig.~\ref{fig:disorder}, we plot the residual energy $r$ for various levels of disorder. Even for a disorder level of 20\% of the magnitude of the interaction strengths the result is much better than random guessing. Moreover, already current quantum devices show much lower disorder levels \cite{king2014algorithm, chen2014qubit, debnath2016demonstration}.
\label{sec:noise}

\section{Translating tree-QAOA parameters into a quantum annealing schedule}
\label{sec:annealing}
\begin{figure*}[t!]
    \centering
    \begin{minipage}{0.5\textwidth}
        \centering
        \includegraphics[width=1\textwidth]{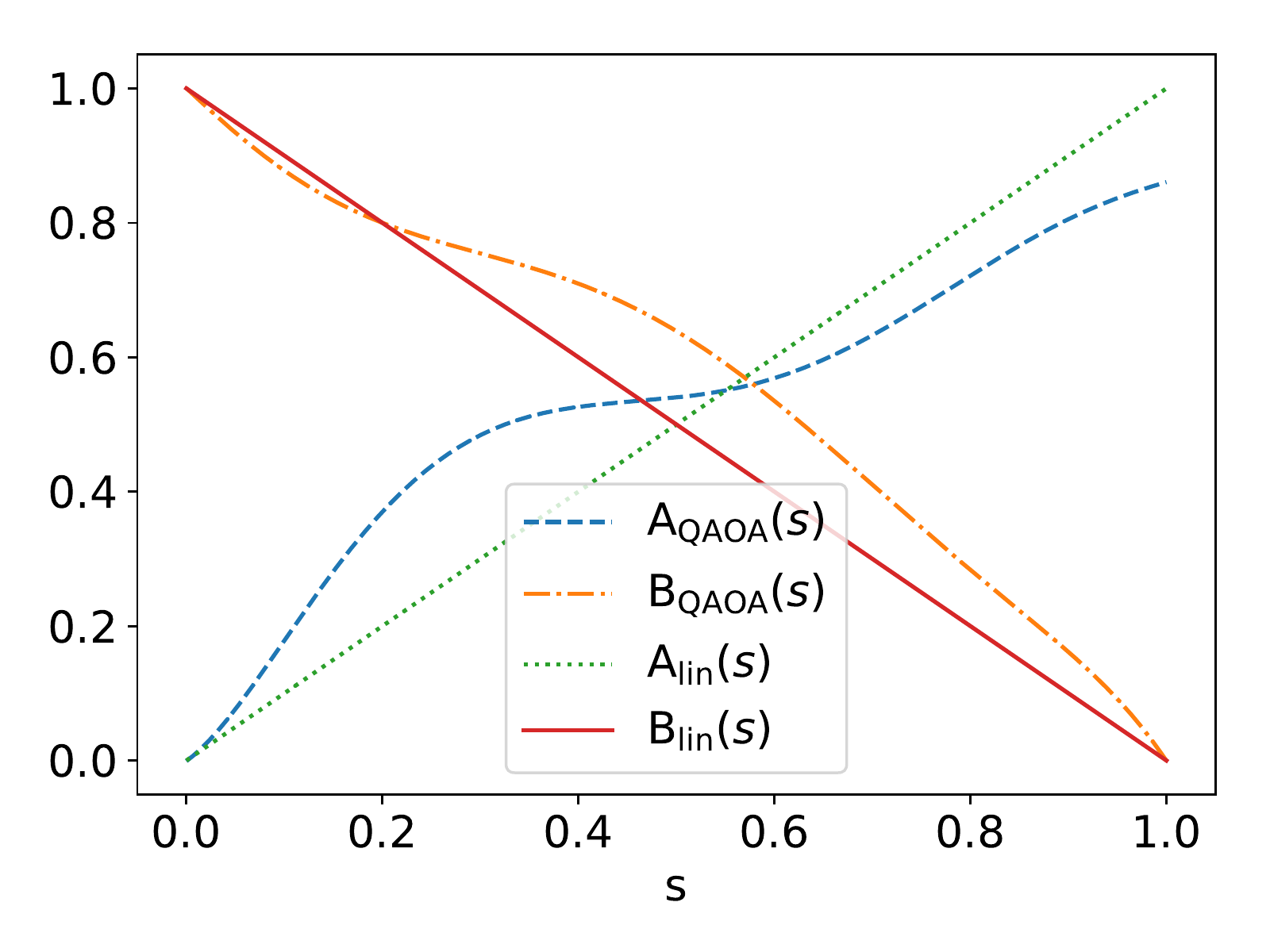}
    \end{minipage}\hfill
    \begin{minipage}{0.5\textwidth}
        \centering
        \includegraphics[width=1\textwidth]{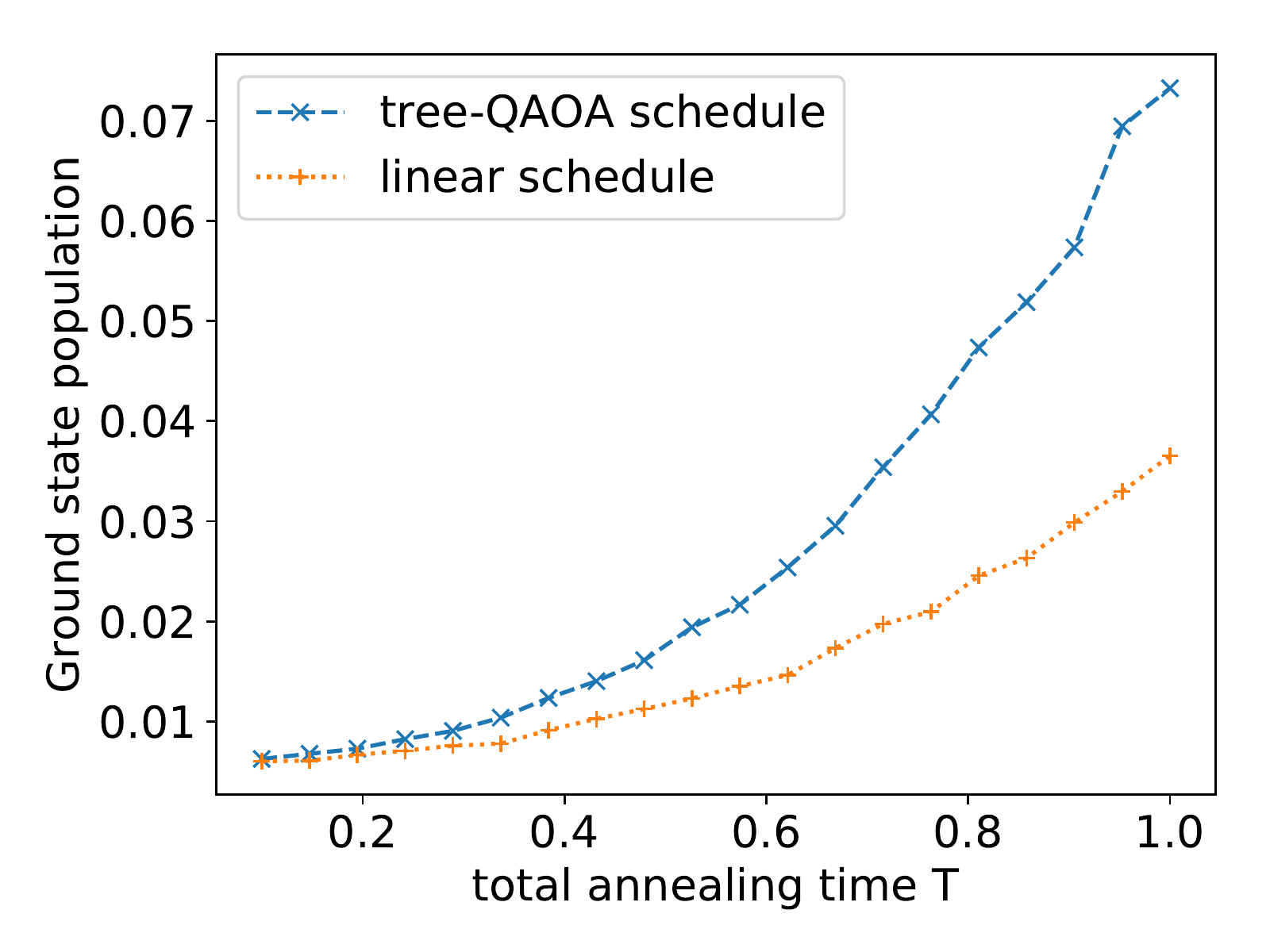}
    \end{minipage}
\caption{(a) A trivial (linear) annealing schedule in comparison to an annealing schedule induced by the tree-QAOA parameters for regular Max-Cut on 3-regular graphs. (b) The ground state population at the end of the annealing run in dependence of the total annealing time $T$ averaged over 100 randomly drawn instances, shown for both the linear ramp and the tree-QAOA annealing schedule.}
\label{fig:annealing}
\end{figure*}

The similarities between QAOA and Quantum Annealing (QA) lead to the question whether good QAOA parameters could also be used to produce good annealing schedules. 

As for QAOA, the objective of QA is to find low-lying energy states of classical problem Hamiltonians $H_\mathrm{P}$. In contrast to the stroboscopic-like application of mixing and problem Hamiltonian in QAOA, in QA, the system starts in the ground state of the mixing Hamiltonian $H_\mathrm{M}$ that is steadily transformed into the problem Hamiltonian $H_\mathrm{P}$ during the annealing process. If this evolution happens slowly enough, the system will stay in the ground state of the instantaneous Hamiltonian
\begin{align}
	H(s)=A(s)H_\mathrm{M}+B(s)H_\mathrm{P},
\end{align}
where $A(s)$ and $B(s)$ describe the ramping functions for the mixing Hamiltonian and problem Hamiltonian respectively.  The parameter $s\in [0,1]$ describes the normalized time with respect to the total annealing time $T$. The probability of staying in the ground state is connected to the minimal energy gap $E_\mathrm{gap}$ of the time-dependent spectrum. To avoid diabatic transitions and to stay in the ground state, the total annealing time has to be longer than $1/E_\mathrm{gap}^2$. Mechanisms, such as slowing down the annealing process at the avoided crossing could be used to avoid transitions. However, such techniques require knowledge about the exact positions of the avoided crossings and therefore knowledge about the full time-dependent energy spectrum. However, as for QAOA, there might exist annealing schedules which are good choices for a broad class of problems. 

To translate the tree-QAOA parameters to an annealing schedule we fit a polynomial of degree $6$ to the absolute values of the parameters found in Sec.~\ref{sec:maxcut}.
We then simulate the outcome of quantum annealing with QuTiP \cite{johansson2013qutip} for the example of Max-Cut on 3-regular graphs with the found tree-QAOA schedule and a trivial annealing schedule, where both ramping functions are linear, i.e. $A_\mathrm{lin}(s)=1-s$ and $B_\mathrm{lin}(s)=s$, see Fig.~\ref{fig:annealing}(a). Moreover, for a fair comparison we normalized the energy scale to $[0,1]$ for both annealing schedules, resulting in the functions shown in Fig.~\ref{fig:annealing}(a).
In Fig.~\ref{fig:annealing}(b), we plot the ground state population averaged over $N=100$ 10-spin-Max-Cut instances for both annealing schedules in dependence of the total annealing time. The annealing schedule induced by tree-QAOA performs significantly better than the linear annealing function. This first observation gives reason to hope that finding good annealing schedules on tree-like structures could improve quantum annealing results in the future. 
\section{Conclusion and outlook}
\label{sec:conclusion}
In this paper, we have introduced a new strategy for inferring control parameters of QAOA. The main advantage of this method is that it does not rely on repetitive calls of the QPU for making parameter updates, but can be simulated with Tensor Networks efficiently on classical hardware and thus embodies a first version of QAOA that does not rely on the QPU for finding the optimal parameters. We studied the performance on instances from different problem classes including up to $N=20$ qubits and compared it to vanilla QAOA setups where classical outer learning loops were used to find good variational parameters. In all situations, our results were either comparable or even better than QAOA with training. Due to the construction of our method, we expect that the advantage will be more pronounced for large problem instances. However, this remains an open question and has to be proven experimentally. We also showed that our method is resilient under the influence of a reasonable amount of disorder on the control parameters of the quantum algorithm, which makes it perfectly suitable for NISQ devices. In addition, translating the tree-QAOA parameters to a annealing schedule lead to better performance of quantum annealing in comparison to a trivial annealing schedule. 

In this work, we solely focused on the performance of randomly generated instances, which is important for real applications. However, an interesting but open question is how this method performs on the hardest instances of a problem class. Moreover, for the next generations of QPUs the noise level will decrease, therefore the realizable circuit depth will increase. To use this method for larger $p$ in the future either approximation methods for contracting the tensor network could be studied or more sophisticated contraction orderings could be found. Furthermore, our method could be adapted to improve classical method such as imaginary time evolution methods \cite{beach2019making} and therefore widen its scope beyond solving combinatorial optimization problems with quantum algorithms. 

\ack
The authors would like to thank Ryan Babbush, Andrea Skolik, Sheir Yarkoni, Florian Neukart and Michael J. Hartmann for fruitful and enlightening discussions. We thank VW Group CIO Martin Hofmann, who enables our research. This project has received funding from the European Union’s Horizon 2020 research and innovation programme under the Grant Agreement No. 828826.
\newline
\bibliographystyle{unsrt}
\bibliography{cc}

\end{document}